\documentclass[%
 reprint,
superscriptaddress,
 aps,
prl,
]{revtex4-2}

\usepackage{chemformula} 
\usepackage[T1]{fontenc}
\usepackage{amssymb}
\usepackage{textcomp}
\usepackage{hyperref}

\newcommand{\mafo}{\mathrm{MgAl_{0.5}Fe_{1.5}O_4}}

\begin{document}
\title{Anisotropic Magnon Spin Transport in Ultra-thin Spinel Ferrite Thin Films -- Evidence for Anisotropy in Exchange Stiffness}

\author{Ruofan Li}
\affiliation
{Laboratory of Atomic and Solid State Physics, Cornell University, Ithaca, NY 14853, USA}

\author{Peng Li}
\affiliation
{Geballe Laboratory for Advanced Materials, Stanford University, Stanford, CA 94305, USA}

\author{Di Yi}
\affiliation
{Geballe Laboratory for Advanced Materials, Stanford University, Stanford, CA 94305, USA}
\affiliation
{State Key Lab of New Ceramics and Fine Processing, School of Materials Science and Engineering, Tsinghua University, Beijing 100084, China}

\author{Lauren Riddiford}
\affiliation
{Geballe Laboratory for Advanced Materials, Stanford University, Stanford, CA 94305, USA}
\affiliation
{Department of Applied Physics, Stanford University, Stanford, CA, 94305, USA}

\author{Yahong Chai}
\affiliation
{School of Integrated Circuits and Beijing National Research Center for Information Science and Technology (BNRist), Tsinghua University, Beijing 100084, China}
\affiliation
{Institute of Microelectronics, Tsinghua University, Beijing 100084, China}

\author{Yuri Suzuki}
\affiliation
{Geballe Laboratory for Advanced Materials, Stanford University, Stanford, CA 94305, USA}
\affiliation
{Department of Applied Physics, Stanford University, Stanford, CA, 94305, USA}

\author{Daniel C. Ralph}
\email{dcr14@cornell.edu}
\affiliation
{Laboratory of Atomic and Solid-State Physics, Cornell University, Ithaca, NY 14853, USA}
\affiliation
{Kavli Institute at Cornell for Nanoscale Science, Ithaca, NY 14853, USA}

\author{Tianxiang Nan}
\email{nantianxiang@mail.tsinghua.edu.cn}
\affiliation
{School of Integrated Circuits and Beijing National Research Center for Information Science and Technology (BNRist), Tsinghua University, Beijing 100084, China}
\affiliation
{Institute of Microelectronics, Tsinghua University, Beijing 100084, China}
\affiliation
{Laboratory of Atomic and Solid-State Physics, Cornell University, Ithaca, NY 14853, USA}



\begin{abstract}
We report measurements of magnon spin transport in a spinel ferrite, magnesium aluminum ferrite $\mafo$ (MAFO), which has a substantial in-plane four-fold magnetic anisotropy. We observe spin diffusion lengths $> 0.8$ $\mathrm{\mu m}$ at room temperature in 6 nm films, with spin diffusion length 30\% longer along the easy axes compared to the hard axes. The sign of this difference is opposite to the effects just of anisotropy in the magnetic energy for a uniform magnetic state. We suggest instead that accounting for anisotropy in exchange stiffness is necessary to explain these results.
 
\end{abstract}

\maketitle



Using magnons, the quanta of spin waves, for information transmission offers the potential for low energy dissipation compared to traditional electronic transport\citep{Chumak2015}. Magnon spin transport has been demonstrated experimentally in both insulating ferrimagnets \citep{Adam1988,Cornelissen2015,Giles2015,Shan2016,Collet2017,Shan2017,Qin2018,Liu2018,J.Liu2018,Oyanagi2020,Demidov2020,Goennenwein2015,ganzhorn2017nonlocal} and antiferromagnets \citep{Lebrun2018,Xing2019,Han2020}. The magnon spin diffusion length, the characteristic propagation length, has been studied  under various conditions of temperature \citep{Cornelissen2016Temp,Ganzhorn2017,Gomez-Perez2020,Oyanagi2020}, magnon chemical potential \citep{Cornelissen2016ChemialPotential,Cornelissen2018,Das2020}, and external magnetic field \citep{Cornelissen2016,Gomez-Perez2020,Oyanagi2020}. Previous measurements on ferrimagnetic insulators have focused on yttrium iron garnet (YIG), either with thick films (>100 nm) which are fully relaxed relative to the substrate  \citep{Cornelissen2015,Shan2016,Gomez-Perez2020} or with thinner films \citep{liu2020}.  In either case,  YIG has very weak magnetic anisotropy, and no anisotropies have been observed in the spin transport.

Here we report measurements of magnon transport in a low-loss spinel material that has been recently stabilized in epitaxial thin-film form, magnesium aluminum ferrite $\mafo$ (MAFO) \citep{Emori2018,Wisser2019,Wisser2019PRA,Wisser2020,Riddiford2019}. When grown epitaxially on an $\mathrm{MgAl_2O_4}$ substrate, MAFO possesses a substantial in-plane cubic anisotropy ($\sim 13$ mT with easy axes in the <110> directions), while maintaining low magnetic damping into the regime of ultrathin films.  We report that magnon spin transport depends strongly on the propagation direction relative to the anisotropy axes, with a spin diffusion length 30\% greater for magnon propagation along the easy axes compared to the hard axes. We argue that this difference has the wrong sign to be explained taking into account only the usual magnetic anisotropy energy which applies to spatially-uniform states, but require also a consideration of anisotropy in the exchange stiffness for nonuniform states. Our results suggest that spin transport measurements can be used as a sensitive probe of the exchange stiffness and that manipulation of this stiffness (e.g., via strain) provides an alternative strategy for controlling magnon spin diffusion.

\begin{figure}[!h]
  \centering
    \includegraphics[width=0.48\textwidth]{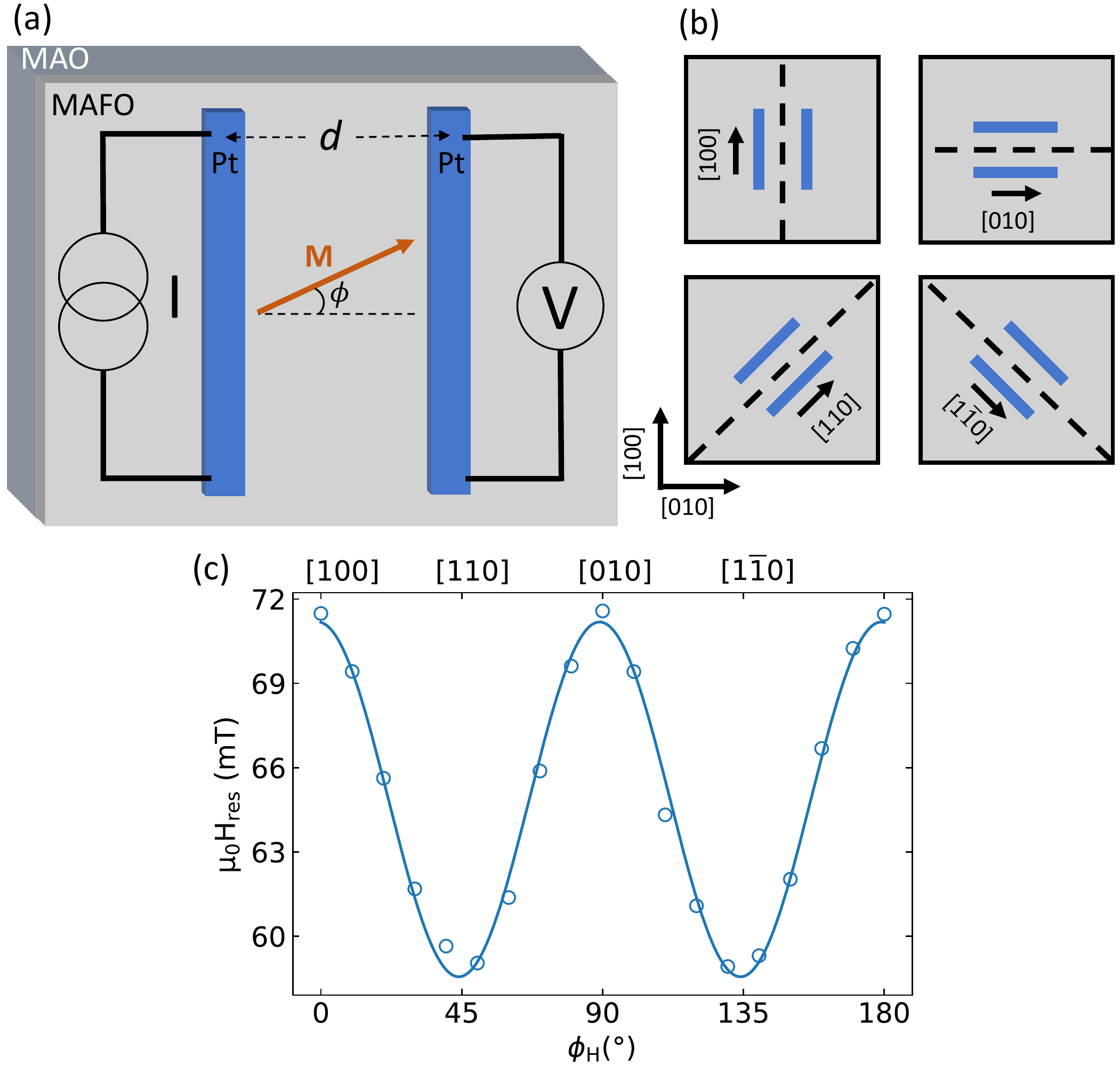}
  \caption{Measurement geometry and magnetic anisotropy of MAFO (a) Schematic layout of the experimental setup (not to scale). (b) Schematics for measuring magnon transport along different directions relative to the crystal axes of the MAFO film (c) FMR resonance field at 9 GHz as a function of the angle ($\phi_\mathrm{H}$) between applied in-plane magnetic field and the [100] crystal axis for a MAFO thin film of 5 nm thickness grown under the same conditions as the 6 nm film. The solid line represents a $\cos (4 \phi_\mathrm{H})$ fit to the measured data.}
  \label{fig:geometry_fmr}
\end{figure}

We employ a measurement geometry commonly used for measuring magnon spin transport -- parallel Pt wires with different separation distances deposited on top of the magnetic insulator to be investigated (Fig.~\ref{fig:geometry_fmr}(a)) \citep{Cornelissen2015,J.Liu2018,Shan2017,Shan2016,Oyanagi2020,Lebrun2018,Han2020}. The Pt wires have widths 200 nm, lengths 10 $\mu$m, and spacings that range from 0.4 to 3.2 $\mu$m.  A charge current passing through one of the Pt wires results in the excitation of magnons in the magnetic film below the wire by the spin Hall effect (SHE) \citep{Vila2007} and, because of a thermal gradient arising from Joule heating, also by the spin Seebeck effect (SSE) \citep{uchida2008}. The excited magnons diffuse through the film to the other Pt wire where they are detected by means of a voltage signal generated by the inverse spin Hall effect (ISHE) \citep{Saitoh2006}.
To measure magnon spin transport in different directions relative to the underlying crystalline film we measure separate devices on the same chip with different orientations of the Pt wires (Fig.~\ref{fig:geometry_fmr}(b)).  

The magnetic insulators we probe are (001)-oriented 6-nm-thick MAFO thin films that are epitaxially grown on $\mathrm{MgAl_2O_4}$ (MAO) substrates \citep{Emori2018} (see Supplemental Material \citep{suppl} for sample growth and fabrication). 
Tetragonal distortion due to epitaxial strain acting on MAFO (3\% lattice mismatch) induces an in-plane 4-fold magnetic anisotropy with an effective field strength of 13 mT as shown by ferromagnetic resonance (FMR) measurements on a MAFO film of similar thickness (5 nm) grown under the same conditions (Fig.~\ref{fig:geometry_fmr}(c)). The angular dependence of the anisotropy field is consistent with cubic symmetry, and the easy axes are along <110>.  The Gilbert damping in MAFO as measured by FMR remains small and isotropic with respect to the angle of in-plane magnetic field relative to the crystal axes \citep{Emori2018}.

For the magnon spin-transport measurements, we pass a low-frequency current (5.9 Hz) through one Pt wire, exciting magnons in the MAFO film via both the SHE and SSE.  All the measurements were performed at room temperature, using different angles and magnitudes of in-plane applied magnetic field, and different spacings between the injector and detector Pt wires. The component of the nonlocal voltage ($V_\mathrm{NL}$) detected in the distant Pt wire that originates from SHE ($V_\mathrm{SHE}$) has a linear dependence on the current ($I$), while the component from SSE ($V_\mathrm{SSE}$), which arises due to a temperature gradient from Joule heating, varies quadratically with $I$. These two kinds of nonlocal voltages can therefore be distinguished by detecting the first ($V_\mathrm{1\omega}=R_\mathrm{1\omega} I$) and second-harmonic ($V_\mathrm{2\omega}=R_\mathrm{2\omega} I^2$) responses using lock-in amplifiers. Depending on their origin, the nonlocal resistances can be then written as
\begin{equation}
R_{1\omega}= R_{SHE}+R_{0,1\omega}
\label{r1w}
\end{equation}
\begin{equation}
R_{2\omega}= R_{SSE}+R_{0,2\omega}
\label{r2w}
\end{equation}
where $R_\mathrm{SHE}$ and $R_\mathrm{SSE}$ represent the nonlocal resistances that we wish to measure arising from the SHE and SSE, while $R_\mathrm{0,1\omega}$ and $R_\mathrm{0,2\omega}$ are offset resistances due to inductive and capacitative couplings in the sample and the measurement setup.

\begin{figure}[!htbp]
  \centering
    \includegraphics[width=1\columnwidth]{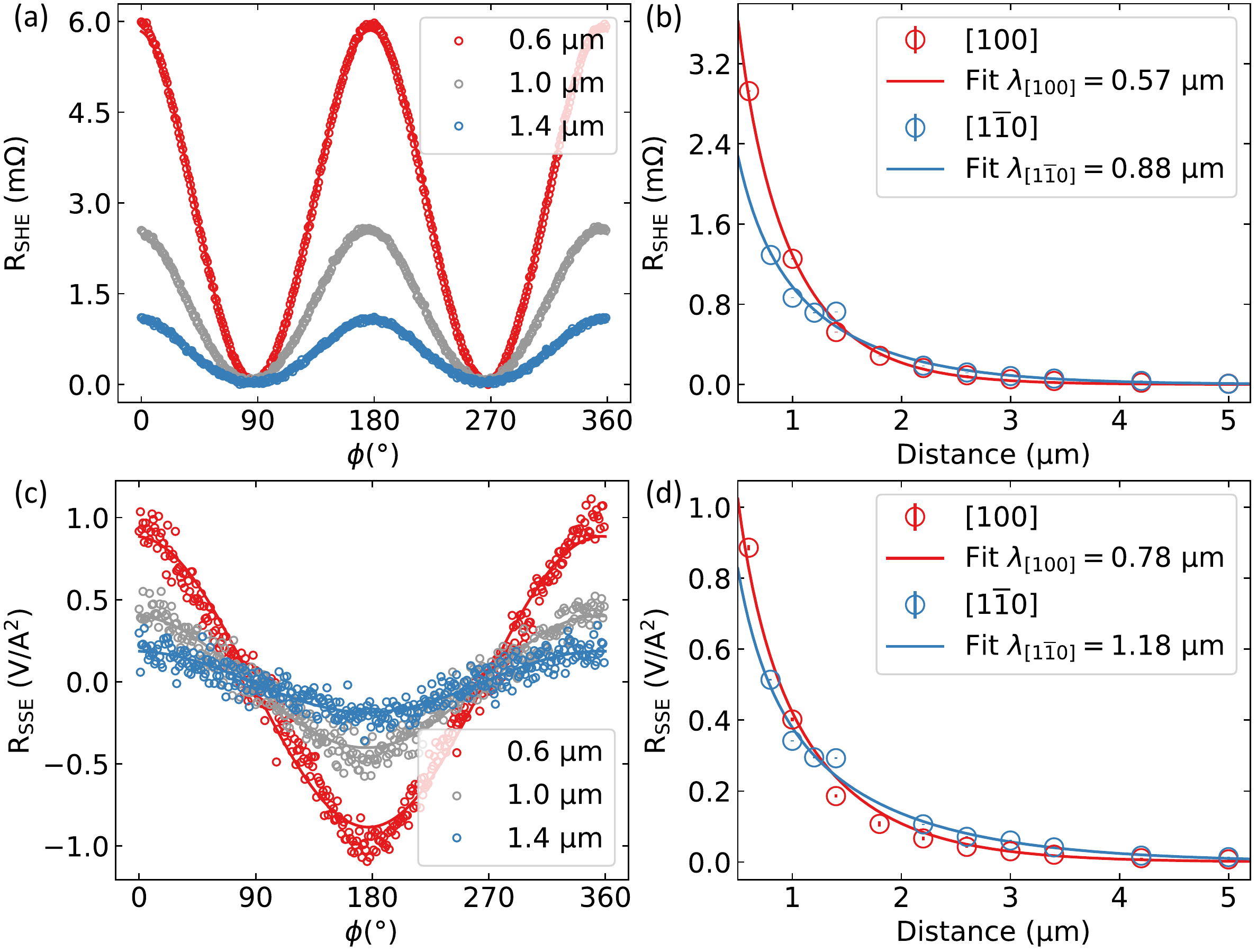}
  \caption{(a) First-harmonic $R_{SHE}$ and (c) second-harmonic $R_{SSE}$ nonlocal signals as functions of the magnetic-field angle $\phi$ for a field magnitude of 75 mT and samples with $d$ = 0.6, 1, and 1.4 $\mathrm{\mu m}$, with the Pt wires oriented along the [100] direction. Solid lines are the fits to $\mathrm{cos^2(\phi)}$ and $\mathrm{cos(\phi)}$ dependence for first and second-harmonic signals respectively. Magnitude of (b) first-harmonic and (d) second-harmonic nonlocal signals as functions of distance between the Pt wires oriented along the [100] and $\mathrm{[1\bar{1}0]}$ directions. The solid lines are fits to Eq.~\ref{decay}.}
  \label{fig:angular_dep_decay}
\end{figure}

To subtract out the constant-impedance parts ($R_{0,1\omega}$ and $R_{0,2\omega}$), we collected $V_\mathrm{1\omega}$ and $V_\mathrm{2\omega}$ as a function of the in-plane magnetic angle $\mathrm{\phi}$ for a field magnitude of 75 mT, as shown in Fig.~\ref{fig:angular_dep_decay}(a). $\mathrm{\phi}$ is defined as the complement of the angle between the magnetization of MAFO and the applied current axis  (see Fig.~\ref{fig:geometry_fmr}(a)). For the electrically-generated nonlocal signal $V_\mathrm{SHE}$, both the injection and the detection of the magnons have a $\mathrm{cos(\phi)}$ dependence coming from SHE and ISHE respectively, resulting in a total dependence approximately $\propto \mathrm{cos^2(\phi)}$  (Fig.~\ref{fig:angular_dep_decay}(a)). In the case of the thermally-generated nonlocal signal $V_\mathrm{SSE}$, the magnon injection is generated by Joule heating, which has no angular dependence, while the detection of the magnons through ISHE varies with angle $\mathrm{\phi}$, which gives rise to an approximately $\mathrm{cos(\phi)}$ angular dependence (Fig.~\ref{fig:angular_dep_decay}(c)).

Although generally magnon conductance decreases with decreasing thickness in magnetic insulator films due to increased damping \citep{Jungfleisch2015}, nevertheless even in 6 nm of MAFO we observe long-range magnon transport across 3.2 $\mu$m gaps. 
The spin diffusion length can be extracted from the decay of $R_\mathrm{SHE}$ and $R_\mathrm{SSE}$ as a function of the separation ($d$) between the Pt wires (injector and detector). This decay can be well fitted to a magnon diffusion model \citep{Cornelissen2015}: 
\begin{equation}
R_{NL}= \frac{C}{\lambda} \frac{\exp{(d/\lambda)}}{1-\exp{(2d/\lambda)}}
\label{decay}
\end{equation}
where $R_{NL}$ could be either $R_\mathrm{SHE}$ and $R_\mathrm{SSE}$, $C$ is a distance-independent constant, and $\lambda$ is an effective spin diffusion length in the direction perpendicular to the Pt wires.

\begin{figure}[!htbp]
  \centering
    \includegraphics[width=0.46\textwidth]{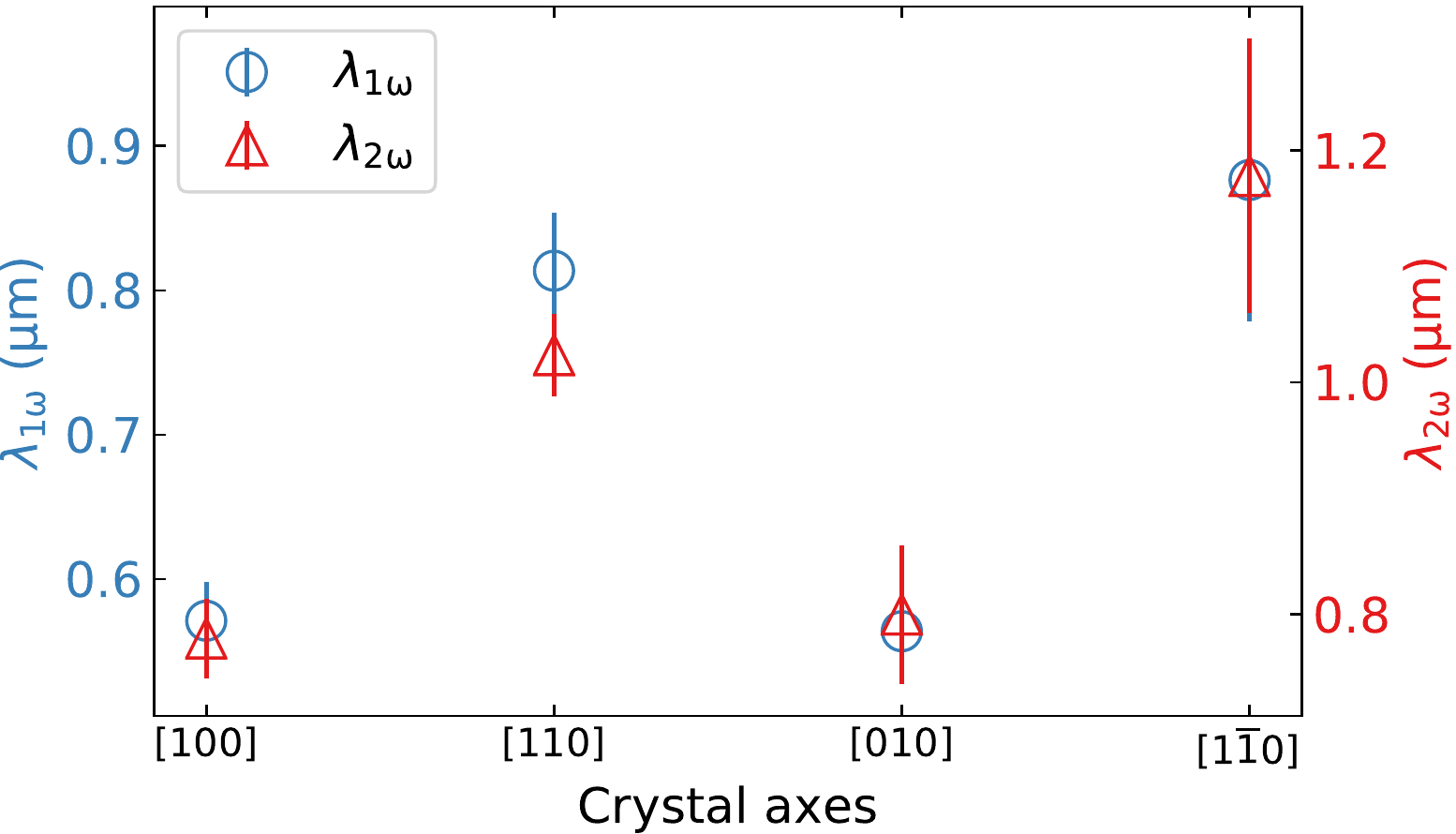}
  \caption{Spin diffusion lengths extracted from the decay of the first (blue circles) and second-harmonic (red triangles) nonlocal signals for wires along corresponding axes. Error bars represent the standard deviations of the fits.}
  \label{fig:lambda_measurement}
\end{figure}

We have investigated the effects of anisotropy by comparing spin diffusion lengths for the orientation of the wires along [100], [110], [010] and $\mathrm{[1\bar{1}0]}$ axes. Fig.~\ref{fig:angular_dep_decay}(b) shows the first-harmonic non-local resistances for the [100] and $\mathrm{[1\bar{1}0]}$ wire orientations plotted as a function of the wire spacing. The dots in the plots correspond to the experimental data while the solid lines show the fits to Eq.~(\ref{decay}). A similar decay of the nonlocal resistance vs.\ spacing was also observed for the second-harmonic signal as shown in Fig.~\ref{fig:angular_dep_decay}(d). The spin diffusion lengths, $\lambda_\mathrm{1\omega}$ and $\lambda_\mathrm{2\omega}$, extracted from the fits for first and second-harmonic signals are shown in Fig.~\ref{fig:lambda_measurement} for the different crystal-axis orientations. 

For both $\lambda_\mathrm{1\omega}$ and $\lambda_\mathrm{2\omega}$ we observe significantly larger (> 30\%) spin diffusion lengths along the <110> family of axes (easy axes) compared to the <100>  axes (hard axes). We also find that the extracted values of $\lambda_\mathrm{2\omega}$ are slightly larger than $\lambda_\mathrm{1\omega}$, which has also been observed previously for YIG thin films \citep{Gomez-Perez2020}. This difference can be explained as due to the different mechanisms by which the nonequilibrium magnon distributions are generated for the two signals. Furthermore, due to a lateral thermal gradient near the Pt injector bar, the second-harmonic voltage can have contributions from both local and nonlocal SSE signals, while for the first-harmonic signal the SHE excites the magnons only locally \citep{Shan2016}.  The angular dependencies of $\lambda_\mathrm{1\omega}$ and $\lambda_\mathrm{2\omega}$ correspond to the same 4-fold symmetry as the in-plane magnetic anisotropy, consistent with the cubic symmetry of MAFO (Fig.~\ref{fig:geometry_fmr}(c)). 

\begin{figure}[!htbp]
  \centering
    \includegraphics[width=0.48\textwidth]{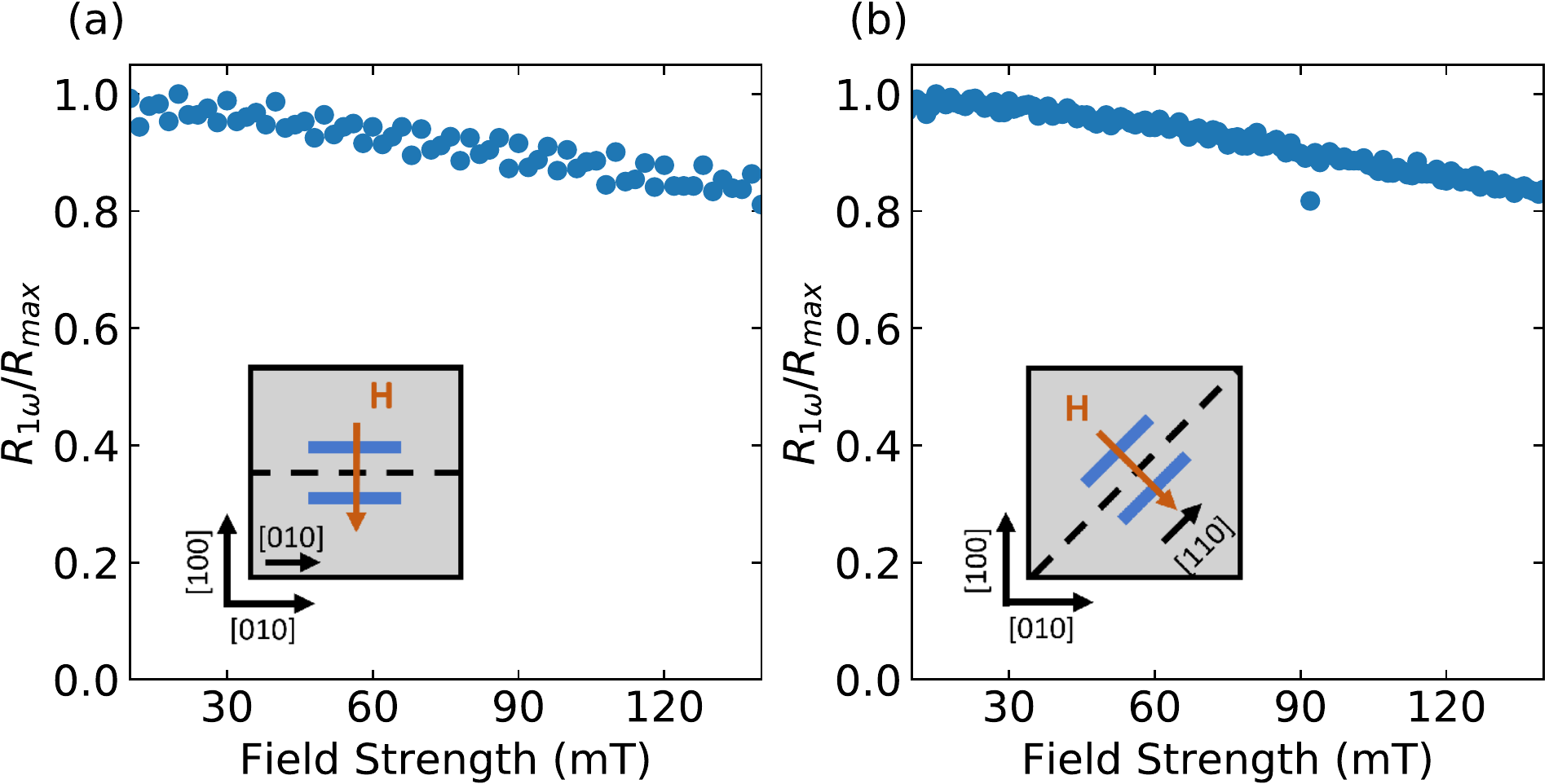}
  \caption{The ratio of first harmonic nonlocal signal with its maximum value as a function of applied magnetic field at $\phi = 0$ for samples with d = 1 $\mathrm{\mu m}$ and Pt wires oriented along (a) hard axis (b) easy axis. The inset in each plot shows the orientation of the Pt wires and applied magnetic field with respect to the crystal axis.}
  \label{fig:field_dependence}
\end{figure}

If one assumes that the primary cause of anisotropic magnon transport in MAFO is simply the anisotropy in the magnetic energy for a uniform magnetic state, it is surprising that the spin diffusion length is longer in the direction of the magnetic easy axis, rather than the reverse. The magnetic anisotropy energy for a uniform magnetic state should cause the same qualitative behavior as an increased applied magnetic field along the easy axis. Both previous measurements on YIG \citep{Cornelissen2016,Gomez-Perez2020}, and our own measurements on MAFO (Fig.~\ref{fig:field_dependence}) show that the magnitude of the nonlocal spin signal decreases as a function of the increasing magnitude of an applied magnetic field, corresponding to a decreased spin diffusion length with increasing magnetic field. This behavior has been ascribed within the context of the SSE to the influence of the magnetic field increasing the energy of long-wavelength magnons \citep{Cornelissen2016}. Quantitatively, the effect of a magnetic field is also far too weak to explain the scale of the effect that we measure.  Figure \ref{fig:field_dependence} shows that a 140 mT magnetic field decreases the spin signal by only 18\%, indicating that an in-plane cubic anisotropy of 13 mT could not generate the 30\% difference we observe. 

The sign of the effect we observe is also surprising within the usual theoretical framework for modeling the energies and group velocities of long-wavelength magnons.  The only type of anisotropy that is ordinarily considered is the anisotropy energy for a uniform magnetic state, accounted for in terms of an anisotropy field $2\mu_0 H_\text{an}$.  For a 4-fold in-plane magnetic anisotropy, the dispersion curve for long-wavelength magnons (taking into account both exchange and magnetic dipole contributions) takes the form \citep{Kalinikos1990,McMichael2004,Sekiguchi2017}
\begin{equation}\label{eq:dispersioncurve}
	\omega(k,\phi_k,\phi_\text{H})=\frac{g \mu_B}{\hbar} \sqrt{B_1 B_2},
\end{equation}
with
\begin{equation}\label{eq:B1}
	B_1 = B + \mu_0 M_\text{eff}(1-P_k) +Dk^2 + \frac{1}{4}\mu_0 H_\text{an}(3+\cos(4\phi_\text{H}))
\end{equation}
\begin{equation}\label{eq:H2}
	B_2 = B + \mu_0 M_\text{eff}P_k \sin^2(\phi_k) +Dk^2 + \mu_0 H_\text{an}\cos(4\phi_\text{H}),
\end{equation}
and where $k = |\vec{k}|$ is the magnitude of the wavevector,  $\phi_k$ is the angle of the wavevector relative to an easy axis, $\phi_\text{H}$ is the angle of the average magnetization relative to the [100] direction, $B$ = 0.075 T in our angle-dependent measurements, $M_\text{eff}$ is the saturation magnetization, $D$ is the exchange stiffness, and $P_k = 1-[(1-e^{-kd})/(kd)]$ with $d$ as the film thickness.  For our MAFO samples,  $\mu_0 M_\text{eff} = 1.5$ T, $2\mu_0 H_\text{an} = 13$ mT, $d = 6$ nm.  For oxide ferrimagnets, a typical value of the exchange stiffness is $D = 5 \times 10^{-17}$ T m$^2$.  The effect of the anisotropy field is to increase the energy of magnons with small values of $k$ for $\phi_k$ near the easy axis, but to cause little change in the energy of magnons with larger $k$ due to the increasing importance of the exchange and dipole terms.  As a result, the group velocity, $v_g = d\omega(k,\phi_k,\phi_\text{H})/dk$ is always decreased by an increase in the magnetic anisotropy energy.  A larger value of $H_\text{an}$ will also decrease the thermal magnon population. Both effects should decrease the spin diffusion length in the direction of a magnetic easy axis.

We therefore draw the conclusion, based on both the sign and magnitude of the effect we observe, that the anisotropic nonlocal spin signal must be caused by crystalline anisotropies which are different from simply the magnetic anisotropy energy for a uniform magnetic state. We considered whether the scattering time for spin relaxation might depend on the orientation of $\vec{k}$ with respect to the anisotropy axes. But if this were the case, we would expect the scattering to also depend on the orientation of $\vec{k}$ with respect to the an applied magnetic field. We do not observe deviations from the behavior $V_\text{SHE} \propto \mathrm{cos(\phi)}$ and $V_\text{SSE} \propto \cos^2(\phi)$ and therefore conclude that scattering time is not $\vec{k}$ orientation dependent (See Supplemental Material \citep{suppl} for residuals in the angular fits).

We suggest, instead, that the anisotropy of our signal is dominated by anisotropies in the exchange energies associated with the MAFO crystal structure. Instead of assuming an isotropic exchange stiffness as in Eqs.~(\ref{eq:dispersioncurve}-\ref{eq:H2}), we can model the exchange stiffness $D$ as a function of the orientation $\vec{k}$ relative to the crystal axes for the long wavelength spin waves that contribute most to the non-local measurements; more specifically $D$ is larger for $\vec{k}$ along the magnetic easy axis so as to increase group velocity in those directions. The possibility of an anisotropic exchange stiffness has been considered previously \citep{Belashchenko2004,Skomski2005,Heide2008}. For non-relativistic exchange processes, the spin stiffness should not depend on the orientation of the magnetization with respect to the crystal axes ($\phi_\text{H}$), but it can depend on the orientation of the wavevector relative to the crystal axes ($\phi_k$).  This is the symmetry required to explain our results without significant deviations from the observed dependence on the angle of magnetic field ($V_\text{SHE} \propto \cos(\phi)$ and $V_\text{SSE} \propto \cos^2(\phi)$).

The existence of anisotropy in exchange stiffness can also help to explain the differences in the magnitude of the spin transport signal extrapolated to small spacings $d$ between source and detector wires -- the fact that the spin signals in the limit of small $d$ become larger for transport along the hard axis compared to the easy axis. The anisotropies in both the exchange stiffness and the energy of the uniform magnetic state have the sign to increase the energies of long-wavelength magnons with $\vec{k}$ along the easy axes, so the population of those magnons will be decreased relative to magnons with $\vec{k}$ along the hard axes.

We are not aware of previous observations of anisotropy in exchange stiffness by broadband ferromagnetic resonance (FMR) or Brillouin light scattering (BLS), the two most common techniques for making direct measurements of exchange stiffness in thin-film samples \citep{Schreiber1996,Klingler2014,Hamrle2009,Haldar2014}. Broadband FMR measures the exchange stiffness in only one direction ($\vec{k}$ perpendicular to the plane of the thin film) since it requires measuring spin-wave standing waves within the film thickness.  To separate the effects of anisotropy in exchange stiffness from a simple magnetic anisotropy energy using BLS would likely require measurements as a function of the magnitude of $\vec{k}$.

In summary, we have measured magnon-mediated spin transport in epitaxially-grown ultrathin (6 nm) $\mafo$ thin films. The small isotropic Gilbert damping parameter ($\sim$0.0015) of these films, their soft magnetism (in-plane coercive field < 0.5 mT), and low processing temperature ($\sim$450 \textcelsius{}) make MAFO an particularly attractive platform for the study of magnon transport and integrated magnonic devices. Unlike previous studies of YIG samples, tetragonally-strained epitaxial MAFO posseses substantial in-plane cubic magnetic anisotropy.  We find also a strong anisotropy in magnon-mediated spin transport, with spin diffusion lengths 30\% larger along the easy axes as compared to that along the hard axes.  The sign of this effect is opposite to what would be expected due simply to the magnetic anisotropy energy of a uniform magnetic state, so we suggest that the anisotropy in spin transport is dominated instead by anisotropy in exchange stiffness.  An exchange stiffness that is larger for $\vec{k}$ long the magnetic easy axis can explain not only the longer spin diffusion lengths for transport along the easy axes but also larger nonlocal spin signals in the limit of small spacing that we observe for transport along the hard axes.  Nonlocal spin wave transport measurements might therefore serve as a sensitive probe of exchange-stiffness anisotropy in thin-film samples. Since crystalline anisotropies can be tuned by strain, we also suggest that strain-mediated manipulation of exchange stiffness might provide a strategy for modulating spin transport in magnetic thin films. \\

\begin{acknowledgments}
We thank Andrei Slavin and Vasyl Tyberkevych for a critique of an initial draft of this paper, and Satoru Emori, Jiamian Hu, Pu Yu, Dingfu Shao and Evgeny Tsymbal for helpful discussions. Research at Cornell was supported by the Cornell Center for Materials Research with
funding from the NSF MRSEC program (Grant No. DMR-1719875). This work was performed
in part at the Cornell NanoScale Facility, a member of the
National Nanotechnology Coordinated Infrastructure, which
is supported by the NSF (Grant No. NNCI-2025233). Research at Tsinghua was supported by the National Natural
Science Foundation (52073158), and the Beijing Advanced Innovation Center for Future Chip (ICFC). Research at Stanford was funded by the Vannevar Bush Faculty Fellowship of the Department of Defense (contract No. N00014-15-1-0045).  L.J.R. acknowledges  support from  the  Air  Force  Office  of  Scientific  Research (Grant  No. FA  9550-20-1-0293) and an NSF Graduate Research Fellowship.
\end{acknowledgments}


\bibliography{mafo}

\end{document}